\documentclass[pra,aps,showpacs,superscriptaddress,twocolumn]{revtex4}
\usepackage{graphicx}
\usepackage{amsmath}
\usepackage{amsfonts}
\usepackage{amssymb}
\usepackage{epsfig}

\begin{document}
\title{
Entanglement sharing in $E\otimes\epsilon$ Jahn-Teller model in
the presence of a magnetic field}
\author{Giuseppe Liberti}\email{liberti@fis.unical.it}\affiliation{
Dipartimento di Fisica, Universit\`a della Calabria, 87036
Arcavacata di Rende (CS) Italy}
\author{Rosa Letizia Zaffino}\affiliation{ Dipartimento di
Fisica, Universit\`a della Calabria, 87036 Arcavacata di Rende
(CS) Italy} \affiliation{INFN - Gruppo collegato di Cosenza, 87036
Arcavacata di Rende (CS) Italy}
\author{Francesco Piperno}\affiliation{
Dipartimento di Fisica, Universit\`a della Calabria, 87036
Arcavacata di Rende (CS) Italy} \affiliation{INFN - Gruppo
collegato di Cosenza, 87036 Arcavacata di Rende (CS) Italy}
\author{Francesco Plastina}\affiliation{ Dipartimento di Fisica, Universit\`a della
Calabria, 87036 Arcavacata di Rende (CS) Italy} \affiliation{INFN
- Gruppo collegato di Cosenza, 87036 Arcavacata di Rende (CS)
Italy}

\date{\today}

\begin{abstract}
We discuss the ground state entanglement of the $E\otimes\epsilon$
Jahn-Teller model in the presence of a strong transverse magnetic
field as a function of the vibronic coupling strength. A complete
characterization is given of the phenomenon of entanglement
sharing in a system composed by a qubit coupled to two bosonic
modes. Using the residual $I$-tangle, we find that three-partite
entanglement is significantly present in the system in the
parameter region near the bifurcation point of the corresponding
classical model.
\end{abstract}
\bigskip
 \maketitle

\section{Introduction}
Tools developed in the realm of quantum information theory are
increasingly being used to investigate fundamental condensed
matter problems \cite{amico}. In particular, many model-systems
exhibiting quantum phase transitions have been explored, and new
insights in their behavior has been gained by studying the
entanglement \cite{oster}, the block entropy \cite{bloc} and the
fidelity \cite{zana}. In particular, it has been demonstrated in
general that, apart from accidental cancellations, entanglement
measures always become singular near the critical points
 \cite{bologn} (in the thermodynamic limit) and exhibit a scaling
behavior (for finite size systems).

Moreover, entanglement has been shown to display not only the
signatures of the critical behavior corresponding to quantum phase
transitions, but also to signal the presence of bifurcations in
the corresponding semiclassical limit \cite{Hines,emary}. This has
been demonstrated, for example, in some spin-boson models in the
strong coupling regime, including the Dicke \cite{brand} and the
Jahn-Teller models \cite{milburn,libertie}. In fact, in the
collective Dicke model, the two aspects of quantum phase
transition and classical bifurcation have been shown to be related
in the adiabatic limit, in which scaling laws have been recently
derived for the ground state entanglement
 \cite{milburn,vidal,libertif}.

Generically, spin-boson models describe the linear coupling of one
 \cite{Jaynes,Tavis,dicke} or many  \cite{leggett} bosonic
modes (typically, photons or phonons) with  electronic or
pseudo-spin degrees of freedom, usually represented as two level
systems (qubits). These models have been used to explore
environment induced decoherence and have been shown to display
peculiar properties of entanglement  \cite{costi}.

In this paper, we concentrate on one model of this class, the
Jahn-Teller (JT) model  \cite{Longuet,eng}, involving an
electron-nuclei system, in which a doubly degenerate electronic
state (usually denoted as $E$) is coupled to a doubly degenerate
nuclear displacement mode ($\epsilon$), with the two bosonic modes
coupled to  different (orthogonal) spin-directions of the qubit.
This is one of the most investigated problems in molecular physics
for which a variety of interesting quantum properties have been
demonstrated, despite the fact that the corresponding Hamiltonian
is not exactly solvable. Particularly relevant from our point of
view, is Ref.  \cite{milburn}, where ground state entanglement has
been investigated for this model-system  in the presence of a
transverse magnetic field, by making use of an approximate
analytic form of the ground state and of numerical diagonalization
with a truncated basis. There, by studying the von Neumann
entropy, it has been shown that the field forces the coupled
system into a maximally entangled state in the large coupling
limit.

Besides this aspect, the $E\otimes\epsilon$ system is interesting
from many respects. Here we concentrate on its multipartite
structure. Indeed, the model describes a tripartite system with an
Hilbert space structure of the kind $2\otimes\infty\otimes \infty$
for which we are able to discuss the sharing properties of
entanglement in the adiabatic limit.

In general, quantifying three-partite entanglement is an extremely
difficult task. For the case of qubits, the CKW conjecture
 \cite{coffman}, recently demonstrated by Osborne and Verstraete
 \cite{osborne2006,alteri}, offers us the powerful instruments of the
monogamy inequality and the residual tangle, which have been
already  employed to interpret some magnetic behaviors
 \cite{firenze}. Related results concerning the monogamy have been
achieved in Ref.  \cite{adesso}, for the case of continuous
variables. However, no general method has been developed for
hybrid systems; that is, those including both discrete and
continuous variables. These systems are extremely interesting  for
many information theoretic applications, including the
implementation of quantum memories or the possibility of
entanglement concentration and purification  \cite{mauro}. In this
respect, we think it is interesting to study some relevant case,
such as the JT model we face in this paper. In a related work,
Tessier et al.
 \cite{tessier} examined the case of two-atom Tavis-Cumnmings
model, making use of the Osborne formula  \cite{osborne} to obtain
the I-tangle.

With these motivations, this paper explores the sharing structure
of entanglement of the $E\otimes\epsilon$ JT system in the
presence of a strong uniform magnetic field, whose presence has
been shown to give rise to interesting consequences in connection
with the Berry Phase  \cite{bevilaqua}. Our approach is based on
the adiabatic procedure  which has been already applied to the
case of a qubit strongly coupled to a single slow resonator
\cite{libertie}.

The paper is organized as follows: In Sec. II we formulate the
$E\otimes\epsilon$ model in the presence of a magnetic field and
discuss its solution in the adiabatic approximation; in Sec. III
various entanglement measures are evaluated, for which some
analytic approximations are derived in Sec. \ref{asympt}. Finally,
Sec. \ref{sect4} summarizes our main findings.
\section{The $E\otimes \epsilon$ model and its solution in the presence of an external field}
\label{primasec} The standard JT model describes a qubit
interacting with two degenerate harmonic modes (conventionally
labelled $\theta$ and $\epsilon$). The model Hamiltonian in the
presence of an external field is the following
\begin{eqnarray}
   H&=&\frac{\omega}{2}\left(p_\theta^2+p_\epsilon^2
   +q_\theta^2+q_\epsilon^2\right)\sigma_0\nonumber\\
   &+&\lambda\left(q_\theta\sigma_x+q_\epsilon\sigma_y\right)+\Delta\sigma_z
    \label{1r}
\end{eqnarray}
where we have chosen unit such that $\hbar=c=1$. Here $\omega$ is
the natural frequency of the identical oscillators, $\Delta$ is
the strength of the magnetic field (taken orthogonal to the
directions of the couplings) and also represents the qubit
transition frequency, $\lambda$ is the coupling constant,
$\sigma_0=I$, $\sigma_x$, $\sigma_y$ and $\sigma_z$ are the usual
Pauli matrix and $(q_\theta, q_\epsilon)$ are real normal
coordinates of the vibrational modes.

The system is invariant under rotations around the magnetic field
axis and thus there is a conserved operator $\hat{J}_z$, such that
$[H,\hat{J}_z]=0$, and which is given by
\begin{equation}\label{gei}
    \hat{J}_z=\hat{L}_z\sigma_0+\frac{1}{2}\sigma_z
\end{equation}
$L_z$ being the $z$ component of the orbital angular momentum
\begin{equation}\label{dphi1}
    \hat{L}_z=q_\theta p_\epsilon-q_\epsilon p_\theta
\end{equation}
We will take advantage of this symmetry to employ the eigenvalues
of $\hat J_z$ as labels of the energy eigenstates.

The ground state of the Hamiltonian will be found in the
Born-Oppenheimer approximation under the assumption of a fast
qubit, which is easily realized for strong external fields
($\Delta \gg \omega$). The whole procedure can be followed more
plainly by rewriting the Hamiltonian (\ref{1r}) in polar
coordinates as follows
\begin{equation}
   H=\frac{\omega}{2}\left[\left(|\vec{p}|^{\,2}+|\vec{q}|^{\,2}\right)
   \sigma_0+\vec{\Theta}\cdot\vec{\sigma}\right]
    \label{hc}
\end{equation}
with $|\vec{p}|^{2}=p_\theta^2+p_\epsilon^2\,,\quad
|\vec{q}|^{2}=q_\theta^2+q_\epsilon^2\,,\quad
\phi=\arctan{(q_\epsilon/q_\theta)}$.

\noindent Notice that the qubit dynamics is governed by the
effective $\vec q$-parametrized magnetic field
\begin{equation}\label{G}
    \vec{\Theta}=(Lq\cos{\phi},Lq\sin{\phi},D)
\end{equation}
where we have introduced the dimensionless parameters
$D=2\Delta/\omega$ and $L=2\sqrt{2}\lambda/ \omega$.

\noindent In the adiabatic assumption of {\it slow} bosonic modes,
and as a first step in the Born-Oppenheimer procedure, we will
regard $\vec{\Theta}$ as approximately static and solve the qubit
dynamics for fixed $\vec q$.

More formally, we look for a solution of the bi-dimensional
Schr\"{o}dinger equation $H|\psi\rangle=E|\psi\rangle$ written in
terms of qubit $|\chi (\vec{q}\,)\rangle$ and oscillator
$\varphi(\vec{q}\,)$ functions as
\begin{equation}\label{deco}
    |\psi\rangle=\int d^{\,2} q \, |\psi(\vec{q}\,)\rangle
    =\int d^{\,2} q \, \varphi(\vec{q}\,) |\vec{q}\,\rangle \otimes |\chi
    (\vec{q}\,)\rangle
\end{equation}
where $|\chi (\vec{q}\,)\rangle$ are the eigenstates of the
``adiabatic'' equation of the qubit part
\begin{equation}\label{adiaham}
\vec{\Theta}\cdot\vec{\sigma}
|\chi_\pm(\vec{q}\,)\rangle=\pm\Theta(q)|\chi_\pm(\vec{q}\,)\rangle
\,,
\end{equation}
which gives the eigenvalues
\begin{equation}\label{dq}
\Theta(q)=|\vec{\Theta}|=\sqrt{D^2+L^2q^2}\,.
\end{equation}
 The two eigenstates of Eq.
(\ref{adiaham}) are
\begin{eqnarray} |\chi_{-}(\vec{q}\,)\rangle&=&
e^{-i{\frac{\phi}{2}}}a({q})|\uparrow\rangle-e^{i{\frac{\phi}{2}}}b({q})|\downarrow\rangle\label{gsdgen1}
\\
|\chi_{+}(\vec{q}\,)\rangle&=&
e^{-i{\frac{\phi}{2}}}b({q})|\uparrow\rangle+e^{i{\frac{\phi}{2}}}a({q})|\downarrow\rangle\label{gsdgen2}
\end{eqnarray}
where $|\uparrow\rangle$ and $|\downarrow\rangle$ are the $\pm 1$
eigenstates of $\sigma_z$, while
\begin{eqnarray}\label{gsdgen3}
    a({q})&=&\frac{1}{\sqrt{2}}\sqrt{{1-\frac{D}{\Theta(q)}}},\\
    b({q})&=&\frac{1}{\sqrt{2}}\sqrt{{1+\frac{D}{\Theta(q)}}}
\end{eqnarray}
The eigenvalues can be then considered as distortion of the
harmonic potential, so that the oscillators are effectively
subject to the adiabatic potentials $W_{\pm} = q^2\pm \Theta(q)$
when the qubit is in $|\chi_{\pm} \rangle$.

The problem, then, reduces to find the solution of a
bi-dimensional Schr\"odinger equation with $W$ as the potential
energy. This is a difficult task, which can be simplified by
exploiting the rotational symmetry.

Since $\hat J_z$ commutes with $H$ and due to the functional
dependence of the adiabatic qubit eigenstates on the polar angle
$\phi$, we can factorize the oscillator wave function in the form
\begin{equation}\label{facto}
\varphi(q,\phi)=(2\pi)^{-1/2}\varphi_{j}(q)e^{i j \phi}
\end{equation}
where $j=\pm1/2,\pm3/2,\dots$ is the eigenvalue of the operator
$\hat{J}_z$.

From Eqs.(\ref{gsdgen1}-\ref{gsdgen2}) the unitary transformation
that diagonalizes the potential energy matrix is obtained as
\begin{equation}\label{unit}
    U=\left(%
\begin{array}{cc}
  e^{-i{\frac{\phi}{2}}}b({q}) & e^{-i{\frac{\phi}{2}}}a({q}) \\
  e^{i{\frac{\phi}{2}}}a({q}) & -e^{i{\frac{\phi}{2}}}b({q}) \\
\end{array}%
\right)
\end{equation}
The transformed Hamiltonian has the form
\begin{equation}
   \tilde{H}=U^\dag H U=\frac{\omega}{2}\left[\left(|\vec{p}|^{2}+
   |\vec{q}|^{2}\right)\sigma_0+\Theta(q)\sigma_z+\Lambda(\vec{q}\,)\right]
    \label{hctrans}
\end{equation}
where
\begin{equation}\label{lq}
    \Lambda(\vec{q}\,)=U^\dag|\vec{p}|^{2}U+2U^\dag\vec{p}\,U\cdot
    \vec{p}=\Lambda_0\sigma_0+\vec{\Lambda}\cdot\vec{\sigma}
\end{equation}
The components of rotated effective field $\Lambda$ are
\begin{equation}\label{la0}
    \Lambda_0=\frac{1}{4}\left(\frac{1}{q^2}+\frac{L^2D^2}{\Theta^4}\right)
\end{equation}
\begin{equation}\label{lax}
    \Lambda_x=-\frac{L}{q\Theta}\left[\hat{L}_z-\frac{D}{\Theta}
    \left(\frac{1}{2}-\frac{D^2}{\Theta^2}\right)\right]
\end{equation}
\begin{equation}\label{lay}
    \Lambda_y=-\frac{DL}{\Theta^2}\frac{\partial}{\partial q}
\end{equation}
and
\begin{equation}\label{laz}
    \Lambda_z=\left[-\frac{1}{q^2}+\frac{L^2}{\Theta(\Theta+D)}\right]\hat{L}_z
\end{equation}
In the absence of magnetic field (the limit $D\rightarrow0$),
\begin{equation}\label{la}
    \Lambda_0=\frac{1}{4
    q^2}\,,\quad\Lambda_x=-\frac{1}{q^2}\hat{L}_z\,,\quad\Lambda_y=\Lambda_z=0
\end{equation}
and the well-known result for the linear $E\otimes\epsilon$
Jahn-Teller model is recovered,  \cite{sato}, i.e.

\begin{eqnarray}
   \tilde{H}&=&\frac{\omega}{2}\left[-\left(\frac{\partial^2}{\partial q^2}
   +\frac{1}{q}\frac{\partial}{\partial
   q}-q^2\right)\sigma_0+Lq\sigma_z\right.\nonumber\\
   &+&\left.\frac{1}{q^2}\left({\hat{L}_z}\sigma_0-\frac{\sigma_x}{2}\right)^2\right]
    \label{linearJT}
\end{eqnarray}
In the strong coupling limit ($L\gg1$), one can neglect the
off-diagonal (non-adiabatic) terms in this expression, so that the
factorization (\ref{facto}) leads to a second-order equation for
the radial function $\varphi_j(q)$ of two adiabatic potential
energy surfaces (APES)
\begin{equation}
\left[-\frac{d^2}{d q^2}-\frac{1}{q}\frac{\partial}{\partial
   q}+q^2\pm Lq+\frac{j^2}{q^2}-\varepsilon_j\right]\varphi_j(q)=0
\label{hc222JT}
\end{equation}
where the term ${j^2}/{q^2}$ plays the role of the centrifugal
energy. In this case, the ground state is characterized by the
quantum number $j=\pm1/2$ and is thus doubly degenerate.

The off-diagonal non adiabatic terms can be neglected directly in
Eq.(\ref{hctrans}) under the assumption of a strong transverse
magnetic field, i.e. $D\gg 1$. This is the regime we will discuss.
For comparison, in this limit, the Hamiltonian (\ref{hctrans})
becomes
\begin{eqnarray}
   \tilde{H}&=&\frac{\omega}{2}\left[-\left(\frac{\partial^2}{\partial q^2}
   +\frac{1}{q}\frac{\partial}{\partial
   q}-q^2\right)\sigma_0+\Theta\sigma_z\right.\nonumber\\
   &+&\left.\frac{1}{q^2}\left({\hat{L}_z}\sigma_0-\frac{\sigma_z}{2}\right)^2\right]
    \label{had}
\end{eqnarray}

The factorization (\ref{facto}) leads to a different equation for
the radial function $\varphi_j(q)$
\begin{equation}
\left[-\frac{d^2}{d q^2}-\frac{1}{q}\frac{\partial}{\partial
   q}+q^2\pm\Theta+\frac{1}{q^2} \left(j\mp
\frac{1}{2}\right)^2-\varepsilon_j\right]\varphi_j(q)=0
\label{hc222}
\end{equation}
with the result that, in the presence of a magnetic field, the
degeneracy present in the linear JT model is broken.

When $D\gg 1$ the motion will remain on the lowest Adiabatic
Potential Energy Surface (APES) given by $W_-=q^2-\Theta(q)$ and
characterized by the quantum number $j=-1/2$ (notice that this
implies that the centrifugal energy equals zero).

Introducing the dimensionless parameter $\alpha={L^2}/{2 D}$, one
can show that for $\alpha\leq1$, the potential $W_{-}(q)$ is just
a broadened harmonic potential surface with a minimum at $q=0$ and
$W_{-}(0)=-D$. For $\alpha>1$, on the other hand, the coupling of
the oscillator with the qubit splits the lowest APES producing a
double-well potential surface with (a circle of) minima at
\begin{equation}\label{ground}
    q=q_0=\sqrt{\frac{D}{2}\left({\alpha-\frac{1}{\alpha}}\right)} \, ,
\end{equation}
with
\begin{equation}\label{eq0}
 W_{-}(q_0)=-\frac{D}{2}\left(\alpha+\frac{1}{\alpha}\right)\, .
\end{equation}

\begin{figure}
\includegraphics{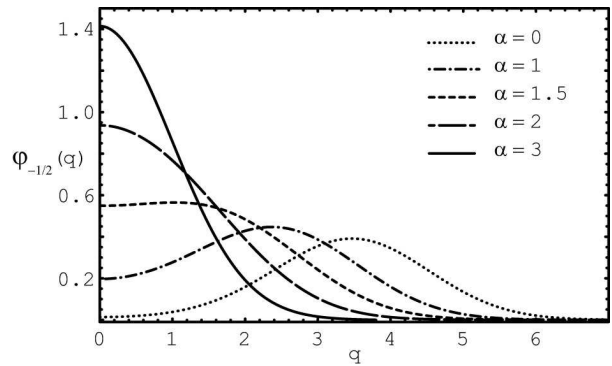}\\
\caption{\label{wfeps} Normalized ground state wave function for
the oscillators in the lower adiabatic potential, for $D=10$ and
different values of $\alpha$.}
\end{figure}

In Fig.(\ref{wfeps}), the ground state wave function
$\varphi_{-1/2}(q)$ is shown for $D=10$ and different values of
$\alpha$. We can see that the maximum probability amplitude is
always found around $q_0$, and that, as $\alpha$ decreases, this
moves far and far away from the origin.
\section{Ground state entanglement}
\label{sectangle} The expression of the ground state obtained in
the previous section enables us to compute  the entanglement
content of the system. We have three independent subsystems: the
qubit, the radial, and the azimuthal  degrees of freedom in which
we have decomposed the two oscillators (from now on, we indicate
these subsystems with the labels $E$, $q$ and $\phi$,
respectively).

In this section we will evaluate the amount of entanglement for
every possible bi-partition and then use the monogamy inequality
to obtain the residual tangle. First, however, we briefly review
the formalism employed.
\subsection{I-Tangle formalism}
To quantify the entanglement for each of the bi-partitions of the
model we will make use of the I-tangle  \cite{rungta}, which for a
rank-2 mixed state $\rho_{AB}$ can be explicitly evaluated as,
 \cite{osborne},
\begin{equation}\label{tauo}
\tau(\rho_{AB})=\mathrm{Tr}(\rho_{AB}\tilde{\rho}_{AB})+
2\lambda_{min}^{(AB)}\left[1-\mathrm{Tr}(\rho_{AB}^2)\right]
\end{equation}
where $\tilde{\rho}_{AB}$ is the result of the action of the
\emph{universal state inverter}  \cite{rungta} on $\rho_{AB}$
\begin{equation}\label{ic}
    \tilde{\rho}_{AB}=S_{A}\otimes S_{B}(\rho_{AB})
\end{equation}
and $\lambda_{min}^{(AB)}$ is the smallest eigenvalue of the M
matrix defined by Osborne  \cite{osborne} which is defined and
then evaluated for our case in appendix \ref{appeosbo}.

The universal inverter $S_i$ is defined to map every pure state
$\rho_i=|\psi\rangle\langle\psi|$ into a positive multiple of its
orthogonal projector, i.e.
$S_i(\rho_i)=\nu_i\left(I-\rho_i\right)$. For an arbitrary
operator $O$, it gives
\begin{equation}\label{so}
    S_i(O)=\nu_i\left [ \mathrm{Tr}(O)I-O\right ]
\end{equation}
where $\nu_i$ is an arbitrary real constant (which we choose to be
unit). The tensor product in Eq.(\ref{ic}), applied to an
arbitrary joint density operator $\rho_{AB}$, is given by
\begin{equation}\label{sptp}
    S_{A}\otimes S_{B} (\rho_{AB}) =
    I_A\otimes I_B-\rho_A\otimes I_B - I_A\otimes\rho_B+\rho_{AB}
\end{equation}
where $\rho_A$ and $\rho_B$ are the reduced density operators
obtained from $\rho_{AB}$. Putting everything together,
\begin{equation}\label{ic2}
    \mathrm{Tr} ( \rho_{AB}\tilde{\rho}_{AB})=
   1-\mathrm{Tr}(\rho_A^2)
    -\mathrm{Tr}(\rho_B^2)+\mathrm{Tr}(\rho_{AB}^2)
\end{equation}
For a joint pure state
($\mathrm{Tr}(\rho_{AB}^2)=\mathrm{Tr}(\rho_{AB})=1$) the I-tangle
(\ref{tauo}) becomes
\begin{equation}\label{ic3}
   \tau_{AB} =2 \left [ 1-\mathrm{Tr}(\rho_A^2)\right ]
\end{equation}
where $\mathrm{Tr}(\rho_A^2)=\mathrm{Tr}(\rho_B^2)$.

We will employ relations (\ref{tauo}) and (\ref{ic3}) several
times in the following.
\subsection{Ground state density operators}
In our case the ground state density operator takes the form
\begin{equation}\label{gsdm0}
    \rho=\int d^2q
    d^2q^\prime\varphi_{-1/2}(\vec{q}\,)\varphi_{-1/2}^\ast(\vec{q\,}^\prime)
    |\vec{q\,}\rangle\langle\vec{q\,}^\prime|
    \left(|\chi_-(\vec{q\,})\rangle\langle\chi_-(\vec{q\,}^\prime)|\right)
\end{equation}
There are six nonequivalent bi-partitions: {(i)} qubit-oscillators
$E\otimes(\phi \, q)$; {(ii)} angular degree of freedom-remainder
$\phi\otimes(E \, q)$; {(iii)} radial degree of freedom-remainder
$q\otimes(E \, \phi)$; {(iv)} angular degree of freedom-qubit
$\phi\otimes E $; {(v)} radial degree of freedom-qubit $q\otimes
E$; {(vi)} radial degree of freedom-angular degree of freedom
$\phi\otimes q$.

To start evaluating the various tangles, it is useful to re-write
the ground state density operator (\ref{gsdm0}) as
\begin{eqnarray}\label{gsdm1}
\rho&=&|a\rangle|f_1\rangle |\uparrow\rangle
\langle\uparrow|\langle f_1|\langle a|
+|b\rangle|f_2\rangle|\downarrow\rangle
\langle\downarrow|\langle f_2|\langle b|\nonumber\\
&-&|a\rangle|f_1\rangle|\uparrow\rangle\langle\downarrow|\langle
f_2|\langle
b|-|b\rangle|f_2\rangle|\downarrow\rangle\langle\uparrow|\langle
f_1|\langle a|\nonumber\\
\end{eqnarray}
where
\begin{eqnarray}\label{ab1}
    |a\rangle &=&\int_0^{\infty} dq \, q \, \varphi_{-1/2}(q) \; a(q) \,
    |q\rangle\,,\\ \label{ab2}
    |b\rangle &=& \int_0^{\infty} dq \, q \, \varphi_{-1/2}(q) \; b(q) \,
    |q\rangle
\end{eqnarray}
are two (non normalized) states of the $q$-mode, while
$|f_i\rangle \, i=1,2$ are the two relevant (and ortho-normal)
states of the angular degree of freedom:
\begin{equation}\label{f12}
    |f_1\rangle=\int_0^{2\pi}\frac{d\phi}{\sqrt{2\pi}}e^{-i\phi}|\phi\rangle\,,
    \,|f_2\rangle=\int_0^{2\pi}\frac{d\phi}{\sqrt{2\pi}}|\phi\rangle
\end{equation}
The situation is similar to that described in Ref.
\cite{milburn}: the angular degree of freedom is constrained to a
two-dimensional subspace of its total Hilbert space and our
tripartite system can be considered as a $2\otimes 2\otimes\infty$
system.
\begin{figure}
\includegraphics{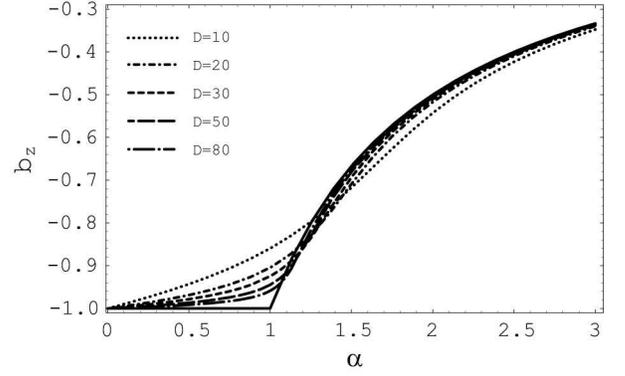}\\
\caption{\label{bzeps} The dependence of the ground state
expectation value $b_z=\langle{\sigma_z}\rangle$ as a function of
the parameter $\alpha$, for various values of $D$. The solid line
corresponds to $D \rightarrow \infty$.}\label{bizeta}
\end{figure}
\begin{figure}
\includegraphics{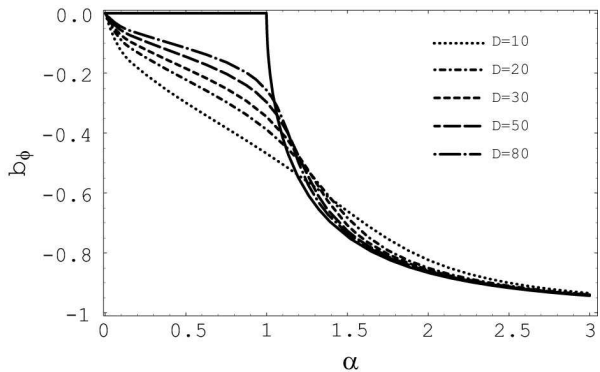}\\
\caption{\label{bLeps} The equatorial component of the Bloch
vector along the $\phi$ direction,
$b_\phi=\langle\cos\phi\,\sigma_x+\sin\phi\,\sigma_y\rangle$ shown
as a function of $\alpha$, for different values of $D$. The solid
line corresponds to $D \rightarrow \infty$.} \label{biphi}
\end{figure}
For the set of states (\ref{ab1}-\ref{ab2}) we have:
\begin{equation}\label{north0}
    \langle a|a\rangle=\frac{1-b_z}{2}\,,\,\langle b|b\rangle=\frac{1+b_z}{2}
\end{equation}
where
\begin{equation}\label{sxm}
b_z=-\int_{0}^\infty q \varphi_{0}^2(q) \frac{D}{\Theta(q)} dq \,
,
\end{equation}
is the $z$-component of the Bloch vector
$\vec{b}=\langle\vec{\sigma}\rangle$ and
\begin{equation}\label{north1}
    \langle a|b\rangle=\langle b|a\rangle=\int_{0}^\infty q
    \varphi_{0}^2(q) \frac{Lq}{\Theta(q)} dq =-{b_\phi}
\end{equation}
where $b_\phi=\langle\cos\phi\,\sigma_x+\sin\phi\,\sigma_y\rangle$
is the equatorial component in the $\phi$ direction.

In Figs.(\ref{bizeta}) and (\ref{biphi}), we show the dependence
of the ground state expectation values $b_z$ and $b_\phi$ on the
dimensionless quantity $\alpha$ for various values of the external
field $D$ (broken lines). The continuous plot describes the case
of very large field ($D\rightarrow \infty$) for which an analytic
expression is obtained in section \ref{asympt}. We will see in the
following that these two parameters completely characterize the
ground state.

From the plots, one can see that for small interaction strengths
(that is, small $\alpha$'s) the external field dominates and
forces the qubit state along its direction; indeed, $b_z\simeq -1$
and $b_{\phi}\simeq 0$. On the other hand, for a large enough
$\alpha$ the qubit is strongly correlated with the angular mode
$\phi$ (loosely speaking, it is `oriented' along $\phi$) with a
small residual polarization along the magnetic field. At
$\alpha=1$ a singular behavior is found for very large fields,
that is analyzed below.

\noindent From eq. (\ref{gsdm1}), the marginal density operators
are easily obtained. For the partitions $\phi\otimes q$ and
$E\otimes q$ one has:
\begin{equation}\label{tmrm}
\rho_{\phi
    q}=\sum_{S=\uparrow,\downarrow}\langle s|\rho|s\rangle=
    |a\rangle|f_1\rangle\langle f_1|\langle a|+
    |b\rangle|f_2\rangle\langle f_2|\langle b|
\end{equation}
\begin{eqnarray}\label{rhor1}
\rho_{Eq}=\sum_{i=1,2}\langle f_i|\rho|f_i\rangle=|a\rangle
|\uparrow\rangle \langle\uparrow|\langle a|
+|b\rangle|\downarrow\rangle \langle\downarrow|\langle b|
\end{eqnarray}
Tracing over $q$ gives a state for $E\otimes \phi$ which has a bit
more involved expression:
\begin{eqnarray}\label{rhosphi0}
\rho_{E \phi}&=&\int_{0}^\infty q\langle q|\rho|q\rangle
dq\nonumber\\
&=&\frac{1+{b_z}}{2}|f_1\rangle |\uparrow\rangle
\langle\uparrow|\langle
f_1|+\frac{1-{b_z}}{2}|f_2\rangle|\downarrow\rangle
\langle\downarrow|\langle f_2|\nonumber\\
&+&\frac{b_\phi}{2}(|f_1\rangle|\uparrow\rangle\langle\downarrow|\langle
f_2|+|f_2\rangle|\downarrow\rangle\langle\uparrow|\langle f_1|)
\end{eqnarray}
As stated above, the reduced density operators are completely
specified by the three set of states introduced above for the
various sub-systems, and by the parameters $b_z$ and $b_{\phi}$.
\subsection{Qubit-oscillators, $\phi$-remainder
and $q$-remainder tangles} In this sub-section, we evaluate the
entanglement of each one of the three subsystems with the
remainder. Since the overall state is pure, the procedure is quite
straightforward. The tangle of the qubit with the two oscillators
is
\begin{equation}\label{tauS}
    \tau_{E(\phi
    q)}=2\left[1-\mathrm{Tr}(\rho_{E}^2)\right] = 1- b_z^2
\end{equation}
The tangle between the angular degree of freedom with the rest of
the system is
\begin{equation}\label{tauphi}
    \tau_{\phi(E q)}=
    2\left[1-\mathrm{Tr}(\rho_{\phi}^2)\right] \, .
\end{equation}
Its expression coincides with $\tau_{E (\phi q)}$ since the
marginal density operator for the $\phi$ degree of freedom has the
same non-zero entries of the qubit one
\begin{equation}\label{rhor3}
\rho_{\phi}= \frac{1+b_z}{2}|f_1\rangle\langle
f_1|+\frac{1-b_z}{2}|f_2\rangle\langle f_2|
\end{equation}
\begin{figure}
\includegraphics{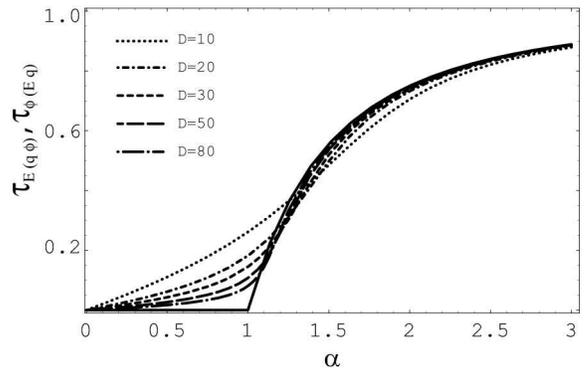}\\
\caption{\label{t1eps} The tangle between the qubit and the
oscillators and between the angular degree of freedom and the rest
as a function of the interaction strength as measured by $\alpha$
for different values of $D$.}
\end{figure}
These two tangles are shown in Fig. (\ref{t1eps}), where it can be
seen that the qubit (as well as the $\phi$ sub-system) essentially
factorizes for small interaction strengths. This is more and more
true for increasing external field and is due to the fact that the
field itself keeps the spin aligned, despite its interaction with
the oscillators. For values of $\alpha$ larger than $1$, the
interaction dominates more and more. This implies that qubit and
angular degree of freedom becomes more and more entangled; indeed,
the tangles saturate to $1$ for large enough $\alpha$'s.

To be more precise, and as better discussed below,  the ground
state contains (for almost every $\alpha$) essentially bi-partite
entanglement as these two degrees of freedom correlate to each
other, with very little involvement of the $q$ part.

To see that this is indeed the case, we start by evaluating the
entanglement to which the radial degree of freedom participates.
The tangle $\tau_{q(E \phi)}$ is given by
\begin{equation}\label{puri2}
    \tau_{q(E \phi)}=
    2\left[1-\mathrm{Tr}(\rho_{q}^2)\right]
\equiv 1-b_z^2-b_\phi^2
\end{equation}
\begin{figure}
\includegraphics{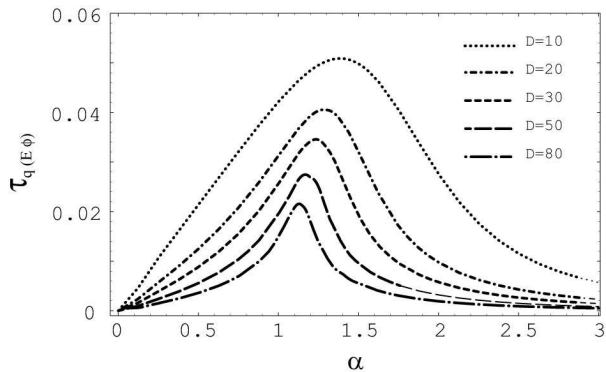}\\
\caption{\label{t2eps} The tangle between between the radial
degree of freedom and the remainder as a function of $\alpha$ for
different values of $D$. Notice that it is notably different from
zero only around $\alpha =1$.}
\end{figure}
This function is shown in Fig. (\ref{t2eps}), where one can see
that the radial degree of freedom is very poorly correlated with
the others. This situation is reminiscent of the one obtained when
a qubit interacts with a single oscillator mode in the presence of
a tilted external field which gives rise to an ``asymmetry'' in
the adiabatic potential, see  \cite{libertie}. The $\phi$-mode,
here, plays exactly the same role of such an asymmetry. In fact,
it destroys the correlations between the radial mode and the qubit
due to the monogamy of entanglement.

It is noteworthy, however, that the entanglement between $E$ and
$q$ is more relevant in the region around $\alpha=1$. Indeed, the
maximum of $\tau_{q (E\phi)}$ moves towards this point as the
field increases and this is exactly the point where, in the strict
adiabatic limit of very large $D$, the tangle becomes
discontinuous.

We show in the following sections that this is precisely the
region in which a true three-partite entanglement (as measured by
the residual tangle) is present. To evaluate the three-partite
correlations, however, we first need to evaluate entanglement for
the other possible bi-partitions in which one of the three
subsystems is traced out. This can be done explicitly thanks to
the Osborne method reviewed above.
\subsection{Angular degree of freedom-qubit tangle}
After tracing over the radial mode $q$, the reduced density
operator for the partition $E\otimes \phi$, eq. (\ref{rhosphi0}),
can be re-written in the form
\begin{equation}\label{rhosphi}
\rho_{E \phi}=\frac{1+\sqrt{b_z^2+b_\phi^2}}{2}|v_1\rangle\langle
v_1|+\frac{1-\sqrt{b_z^2+b_\phi^2}}{2}|v_2\rangle\langle v_2|
\end{equation}
where
\begin{eqnarray}\label{v1}
    |v_1\rangle&=&\beta_1|f_1\rangle|\uparrow\rangle+\beta_2|f_2\rangle|
    \downarrow\rangle \\
    \label{v2}
     |v_2\rangle&=&\gamma_1|f_1\rangle|\uparrow\rangle-
     \gamma_2|f_2\rangle|\downarrow\rangle
\end{eqnarray}
with
\begin{equation}\label{beta1}
    \beta_1=\left[1+\left(\frac{b_z}{b_\phi}+
    \sqrt{1+\frac{b_z^2}{b_\phi^2}}\right)^2\right]^{-1/2}
    \,,\,\beta_2=\sqrt{1-\beta_1^2}
\end{equation}
and
\begin{equation}\label{gamma1}
    \gamma_1=\left[1+\left(-\frac{b_z}{b_\phi}+
    \sqrt{1+\frac{b_z^2}{b_\phi^2}}\right)^2\right]^{-1/2}
    \,,\,\gamma_2=\sqrt{1-\gamma_1^2}
\end{equation}
The vectors $|v_i\rangle, i=1,2$ are the only eigen-kets of
$\rho_{E \phi}$ with non-zero eigenvalues given by
$r_i=\left(1\pm\sqrt{b_z^2+b_\phi^2}\right)/{2}$.

This form (which, by the way, shows that the matrix has rank two)
is particularly useful to apply the Osborne procedure. A
straightforward calculation gives the tangle in the form
\begin{equation}\label{tafhq}
\tau_{E \phi}= \frac{1-b_z^2}{2}\left(1+2\lambda_{min}^{(E
\phi)}\right)+ \frac{b_\phi^2}{2}\left(1-2\lambda_{min}^{(E
\phi)}\right)
\end{equation}
where $\lambda_{min}^{(E \phi)}$ is obtained in appendix
\ref{appeosbo}
\begin{equation}\label{lminsfp}
\lambda_{min}^{(E
\phi)}=\frac{1}{4}\left(1-\sqrt{1+\frac{8b_z^2}{b_z^2+b_\phi^2}}\,
\right)
\end{equation}
\subsection{$q$-$\phi$ and $q$-$E$ tangles}
The two remaining bi-partitions of the system are those consisting
of the radial degree of freedom  and either the angular mode or
the qubit. These turn out to have no entanglement at all. Indeed,
one has:
\begin{eqnarray}\label{tildesfq}
    && \mathrm{Tr}(\rho_{Eq}\tilde{\rho}_{Eq})=
    \mathrm{Tr}(\rho_{\phi q}\tilde{\rho}_{\phi q})=
    \frac{1-b_z^2-b_\phi^2}{2} \\
    && \lambda_{min}^{(E q)}=\lambda_{min}^{(\phi q)}=-
    \frac{1-b_z^2-b_\phi^2}{2(1-b_z^2)} \label{lminsfq}
\end{eqnarray}
Putting everything together in eq. (\ref{tauo}), one has
\begin{equation}\label{tasfq}
\tau_{E q}=\tau_{\phi q}=0
\end{equation}
\begin{figure}
 \includegraphics{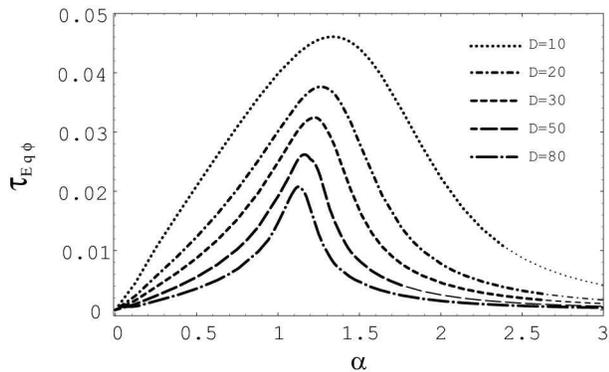}\\
 \caption{\label{t4}The I-residual tangle given in eq.
 (\ref{rt2}), shown for different values of the external magnetic field. }
\end{figure}
\subsection{Residual tangle}
The amount of entanglement for the various bi-partitions that we
have evaluated above, do not give by themselves any indication
neither on the sharing properties nor on the global, three-partite
quantum correlations. Coffman et al.  \cite{coffman} have explored
this problem in a system of three qubits and introduced a quantity
known as the residual tangle, to describe the collective
entanglement content of a state:
\begin{equation}\label{rt0}
    \tau_{ABC}=\tau_{A(BC)}-\tau_{AB}-\tau_{AC}
\end{equation}
When subsystems $A$, $B$ and $C$ are entangled with each other,
the tangle of $A$ with $B$ plus the tangle of $A$ with $C$ cannot
exceed the tangle of $A$ with the joint subsystem $BC$. This
result has been proved valid for any multipartite state of qubits
 \cite{osborne2006}.

In the $E\otimes \epsilon$ JT model, we cannot simply use the
definition (\ref{rt0}) of the residual tangle since our three
subsystems no longer have equal Hilbert space dimension and
symmetry under permutations of the subsystems, which is present in
eq. (\ref{rt0}) would be lost.

Tessier et al.  \cite{tessier} have faced a similar problem and
proposed to generalize the quantity (\ref{rt0}) by just taking the
average of the three residual tangles to introduce the
\emph{I-residual tangle} which has, by definition, the desired
permutation invariance:
\begin{equation}\label{rt}
    \tau_{E \phi q}=\frac{1}{3}\left[\tau_{E\phi q}^{(1)}+
    \tau_{E\phi q}^{(2)}+\tau_{E\phi q}^{(3)}\right]
\end{equation}
where
\begin{eqnarray}\label{rtva}
    \tau_{E\phi q}^{(1)}&=&\tau_{E(\phi q)}-
    \tau_{E \phi}-\tau_{E q}\\
    \tau_{E\phi q}^{(2)}&=&\tau_{\phi(E q)}-
    \tau_{E \phi}-\tau_{\phi q}\\
    \tau_{E \phi q}^{(3)}&=&\tau_{q(E \phi)}-
    \tau_{E q}-\tau_{\phi q}
\end{eqnarray}
In our case, $\tau_{E \phi q}^{(1)}=\tau_{E \phi
q}^{(2)}\neq\tau_{E \phi q}^{(3)}$, and one easily obtains
\begin{equation}\label{rt2}
    \tau_{E\phi q}=\frac{2}{3}\tau_{q(E \phi)}
    \left(1-\lambda_{min}^{(E\phi)}\right)
\end{equation}
This quantity is shown in Fig. (\ref{t4}), from which the
similarity with the plots of Fig. (\ref{t2eps}) can be easily
grasped. This is due to the fact that the $q$ mode is only
involved in genuinely three-partite entanglement as it does not
present any bi-partite quantum correlation neither with the qubit
nor with the angular mode taken alone.

Again, we notice that the residual tangle is present only within a
small region around $\alpha=1$.
\section{Asymptotic behavior of the entanglement}
\label{asympt} In order to obtain an analytic estimation of the
physical quantities evaluated above and for the various
entanglement measures introduced, we would need an expression for
the ground state wave function $\varphi_{-1/2}(q)$. It is possible
to obtain analytically this function under some reasonable
approximation for the effective adiabatic potential. In the
following we report three distinct approximations, valid in the
regimes of {\it i}) small coupling, $\alpha \ll 1$; {\it ii}) very
large coupling $\alpha \gg 1$; and {\it iii}) around the
cross-over value $\alpha \approx 1$.
\subsection{Small coupling regime}
For $\alpha \ll 1$ the adiabatic potential $\Theta$ in the
Schr\"{o}dinger equation (\ref{hc222}) is approximately harmonic,
and the main effect of the qubit is to re-normalize the value of
the oscillator frequency by a factor $k=\sqrt{1-\alpha}$. As a
result, in this regime the adiabatic ground state wave function
for the oscillator is well approximated by the gaussian
\begin{equation}
\varphi_{-1/2}(q) = \left(2 k\right)^{\frac{1}{2}} \, \exp \left
\{ - \frac{k}{2} \, q^2 \right \} \, .
\end{equation}
By repeating the various steps of the previous section, one can
obtain approximate expressions for the various tangles introduced
above, valid to first order in $\alpha$. For example,
\begin{equation}\label{sepD} \tau_{E(\phi q)}\equiv \tau_{\phi(E
q)}\simeq \frac{2\alpha}{D}\,\,,\tau_{E
\phi}\simeq\frac{\pi\alpha}{2D}
\end{equation}
\begin{equation}\label{sepsD}
\tau_{E q \phi} \equiv \tau_{q(E \phi)}\simeq
\left(2-\frac{\pi}{2}\right)\frac{\alpha}{D}
\end{equation}
which we checked to be in very good agreement with the numerical
solution given above, and which describe the start-up of
entanglement as soon as the interaction is switched on. The last
equation shows that (to first order in $\alpha$), the radial mode
is involved only in three-partite entanglement.

\subsection{Strong coupling regime}
For $\alpha \gg 1$, the lowest eigenstate should be localized at
the minimum of the lowest potential surface. Therefore, by
expanding the potential around this minimum [the $q_0$ of
Eq.(\ref{ground})] and by retaining up to second order terms, the
Schr\"{o}dinger equation for the lowest sheet can be viewed as the
equation for a bi-dimensional shifted harmonic oscillator.

Letting $\tilde q=q-q_0$, to be the distance from the minimum, the
approximate adiabatic equation for the ground state with $j=-1/2$,
becomes
\begin{equation}\label{hstrong2}
\left[\frac{d^2}{d \tilde q^2}+\frac{1}{q_0}\left(1-\frac{\tilde
q}{q_0}\right)\frac{d}{d \tilde q}+v_0-\kappa^2 \tilde
q^2+\varepsilon_{-1/2}\right]\varphi_{-1/2}(\tilde q)=0
\end{equation}
where $v_0 = \frac{D}{2 \alpha} (\alpha^2-1)$ is an energy shift,
and $$\kappa= \left (1-\frac{1}{\alpha^2}\right )^{1/2}\simeq 1-
\frac{1}{2\alpha^2}$$ is, again, a renormalization factor for the
oscillator frequency.

To obtain analytic estimates for large $\alpha$, we can take as an
approximate adiabatic ground state for the oscillator the wave
function
\begin{equation}
 \varphi_{-1/2} (\tilde q) \simeq \left(\frac{\kappa q_0^2}{\pi}\right)^{{1}/{4}} \,
\exp \left \{- \frac{\kappa}{2}\tilde q^2 \right \} \, .
\label{due1}\end{equation} In this regime, we have an almost
complete quantum correlation between the qubit and the $\phi$
mode:
\begin{equation}\label{sepcD}
\tau_{E(\phi q)}= \tau_{\phi(E q)}=\tau_{E \phi}\simeq
1-\frac{1}{\alpha^2}\,,
\end{equation}
On the other hand, the $q$ mode is almost factorized since its
wave function is very localized. As a result, the residual tangle
is very close to zero (the leading contribution being of third
order in $1/\alpha$):
\begin{equation}\label{sepscD} \tau_{q(E \phi)}\simeq
\frac{1}{\alpha^3D}\,, \qquad \tau_{E q
\phi}\simeq\frac{2}{3\alpha^3D}
\end{equation}
\subsection{Critical region}
The coupling value corresponding to $\alpha=1$ divides an
essentially separable regime from an entangled one. This point
corresponds to a bifurcation in the appropriate semiclassical
analogue  \cite{milburn}, and we have shown that the region of
parameters around $\alpha=1$ is the only one with non negligible
residual tangle. We have also shown that, when the magnetic field
increases, this cross-over becomes more and more sharp until a
singular behavior is found in the $z$-magnetization and (as a
consequence) in the entanglement measures.

In this section we seek for an analytic description of the system
in this parameter region and show that a scaling behavior is found
with respect to $D$. For this reason we call this a critical
region.

Above, we have defined the adiabatic potential as $$W_-(q) =
q^2-\Theta(q) \equiv q^2 - \sqrt{D^2+L^2q^2} \, .$$ For
$\alpha\sim 1$ it can be approximated with the quartic expression
\begin{equation}\label{quartic}
W_{-}(q)\simeq-D+(1-\alpha)q^2+\frac{\alpha^2}{2D}q^4
\end{equation}
that describes an anharmonic oscillator for $\alpha\leq 1$,
whereas, for $\alpha\geq1$, it is a double-well potential. As in
the single oscillator case  \cite{libertie},  this implies that a
crossover between a localized state and a Schr\"{o}dinger cat-like
state is obtained. This, in turn, implies a drastic change in the
behavior of entanglement.

This approximate potential apparently depends on the two
independent parameters $\alpha$ and $D$, but a reduction to a
single-parametric problem can be obtained with the help of
Symanzik scaling  \cite{simon}. This is done by re-casting the
Schr\"odinger equation (always written for $j=-1/2$, see section
\ref{primasec}), into the equivalent form
\begin{equation}\label{hamscaling}
\left[-\frac{d^2}{dx^2}-\frac{1}{x}\frac{d}{d x}+\zeta
x^2+x^4\right]\varphi_{-1/2}(x;\zeta)
=e_{g}\left(\zeta\right)\varphi_{-1/2}(x;\zeta)
\end{equation}
where $x=q(\alpha^2/2D)^{1/6}$ is a scaled variable. The only
remaining scale parameter is, then,
$\zeta=\left({2D}/{\alpha^2}\right)^{2/3}(1-\alpha)$, while the
ground-state energy is rewritten as
\begin{equation}\label{sr}
\varepsilon_{g}\equiv
\varepsilon_{-1/2}=-D+\left(\frac{\alpha^2}{2D}\right)^{1/3}e_{g}\left(\zeta\right)
\end{equation}

It can be shown that all of the qubits and oscillator expectation
values can be expressed in terms of the diagonal moments:
\begin{equation}\label{genint}
   \langle q^{\nu} \rangle=\int_{0}^\infty q^{\nu+1} \varphi_{-1/2}^2(q) dq=
   \left(\frac{2D}{\alpha^2}\right)^{\nu/6}\langle x^{\nu} \rangle\,,
\end{equation}
where
\begin{equation}\label{genint2}
   \langle x^{\nu} \rangle=\int_{0}^\infty x^{\nu+1} \varphi_{-1/2}^2(x;\zeta) dx
\end{equation}
In fact, the parameter $\zeta$ is very small for $\alpha\approx 1$
and we can obtain analytic approximations for every physical
quantity we need, by retaining only the first orders of their
Taylor expansion in $\zeta$.

For example, the two relevant components of the Bloch vector of
the qubit, taken {\it i}) along the external field ($b_z$), and
{\it ii}) in the equatorial plane along the $\phi$ direction
($b_{\phi}$), have the approximate expressions
\begin{eqnarray}\label{bzas}
   b_z&\simeq&-1+\left(\frac{2\alpha}{D^2}\right)^{1/3}\langle x^{2} \rangle-
   \frac{3}{2}\left(\frac{2\alpha}{D^2}\right)^{2/3}\langle x^{4} \rangle\\\label{bphias}
    b_\phi&\simeq&-\sqrt{2}\left[\left(\frac{2\alpha}{D^2}\right)^{1/6}\langle x \rangle
    -\left(\frac{2\alpha}{D^2}\right)^{1/2}\langle x^{3} \rangle\right]\,.
\end{eqnarray}
These forms for the components of $\vec b$ can be plugged in the
general relations for the various tangles obtained in section
\ref{sectangle}, to get
\begin{equation}\label{puriD}
\tau_{E(\phi q)}= \tau_{\phi(E q)}\simeq
\left(\frac{4}{D}\right)^{2/3}\langle x^2 \rangle\,,
\end{equation}
\begin{equation}\label{purisD}
\tau_{E \phi}\simeq\left(\frac{4}{D}\right)^{2/3}\langle
x\rangle^2
\end{equation}
and
\begin{equation}\label{puri2D}
\tau_{E q \phi} \equiv \tau_{q(E
\phi)}\simeq\left(\frac{4}{D}\right)^{2/3}\left(\langle
x^2\rangle-\langle x\rangle^2\right)
\end{equation}
All the quantities can be evaluated explicitly once we know the
various moments of the scaled position $x$ at $\zeta=0$. These,
however, are just constant numerical values, so that the physical
dependence on $D$ and $\alpha$ can be already read from the
formula above. In particular, a power-law behavior is found, and
both the bi-partite and the residual tangles become singular as
$D^{-2/3}$.

For completeness, we give the numerical values of the first
moments of the scaled position which are involved in the formula
above. For $\alpha= 1$, the problem is reduced to the
bi-dimensional motion in a pure quartic potential
\begin{equation}\label{pqo}
\left(-\frac{d^2}{dx^2}-\frac{1}{x}\frac{d}{d
x}+x^4\right)\varphi_{-1/2}(x;0)=e_{-1/2}(0)\varphi_{-1/2}(x;0)
\end{equation}
whose energy and all of the moments can be computed numerically.
One obtain $e_{-1/2}(0)\simeq 2.3448$, $\langle x \rangle\simeq
0.72737$ and $\langle x^2 \rangle\simeq 0.6515$.

By using these numerical values in Eqs. (\ref{puriD}),
(\ref{purisD}), and (\ref{puri2D}), we obtain that the scaling
with $D$ of the various tangles at $\alpha=1$ is essentially
indistinguishable from the numerical behaviors for large enough
fields (i.e. as long as $D> 10$).
\section{Summary}\label{sect4}
We have discussed the sharing structure of entanglement in
$E\otimes \epsilon$ JT model in the presence of a strong external
field. Using an average residual $I$-tangle obtained from the
monogamy inequality, we have shown that three-partite correlations
are important near the point in parameter space that corresponds
to the bifurcation of the corresponding classical system. This
point divides a separable from an entangled region, and a singular
behavior of entanglement is obtained in the strict adiabatic
limit. By a detailed analysis performed near this point, we have
derived a scaling behavior with respect to the external magnetic
field and identified its ``critical'' exponent.

\appendix
\section{Osborne M matrix} \label{appeosbo}
The central ingredient required for the computation of the
$I$-tangle in Eq.(\ref{tauo}) is the real symmetric $3\times 3$
matrix $M_{ij}$, derived by Osborne in Ref. \cite{osborne}, for a
density operator $\rho$ expressed as a convex combination of its
eigenvectors:
\begin{equation}\label{doa}
    \rho=p|v_1\rangle\langle v_1|+(1-p)|v_2\rangle\langle v_2|
\end{equation}
The independent matrix elements of $M$ are constructed in terms of
the tensor
\begin{eqnarray}\label{gammaij}
    T_{ijkl}&=&\mathrm{Tr}(\gamma_{ij}\tilde{\gamma}_{kl})\nonumber\\
    &=&\mathrm{Tr}(\gamma_{ij})\mathrm{Tr}({\gamma}_{kl})-
    \mathrm{Tr}_A(\mathrm{Tr}_B(\gamma_{ij})\mathrm{Tr}_B({\gamma}_{kl}))
   \nonumber\\&-&
     \mathrm{Tr}_B(\mathrm{Tr}_A(\gamma_{ij})\mathrm{Tr}_A({\gamma}_{kl}))+
      \mathrm{Tr}(\gamma_{ij}{\gamma}_{kl})
\end{eqnarray}
where $\gamma_{ij}=|v_i\rangle\langle v_j|$. For the partition
$E\otimes \phi$ one obtains
\begin{eqnarray}\label{Tijkl}
    T_{1111}&=&4\beta_1^2\beta_2^2\,,\nonumber\\
    T_{1112}&=&T_{1121}=-2\left(\beta_1^3\gamma_1-\beta_2^3\gamma_2\right)\,,\nonumber\\
    T_{1122}&=&T_{2211}=T_{1221}=T_{2112}=1-2\left(\beta_1^2\gamma_1^2+\beta_2^2\gamma_2^2\right)\nonumber\\
    T_{1222}&=&T_{2122}=-2\left(\beta_1\gamma_1^3-\beta_2\gamma_2^3\right)\,,\nonumber\\
    T_{2222}&=&4\gamma_1^2\gamma_2^2\,.
\end{eqnarray}
from which, using Eqs. (\ref{beta1}) and (\ref{gamma1}), we obtain
that the only non-zero matrix elements are
\begin{eqnarray}\label{Tij}
    M_{11}&=&\frac{b_z^2}{b_z^2+b_\phi^2}\,,\nonumber\\
    M_{13}&=&M_{31}=\frac{b_z b_\phi}{b_z^2+b_\phi^2}\,,\nonumber\\
    M_{33}&=&\frac{b_\phi^2-b_z^2}{b_z^2+b_\phi^2}
\end{eqnarray}
The eigenvalues of this M matrix are thus
\begin{equation}\label{lm}
    \lambda_{\pm}^{(E \phi)}=\frac{1}{4}\left(1\pm\sqrt{1+\frac{8b_z^2}{b_z^2+b_\phi^2}}\right)
\end{equation}



\begin{thebibliography}{99}
\bibitem{amico}
L. Amico, R. Fazio, A. Osterloh, and V. Vedral, quant-ph/0703044
(2007).
%
\bibitem{oster}
A. Osterloh, L. Amico, G. Falci, and R. Fazio, Nature {\bf 416},
608 (2002); T. J. Osborne, and M. A. Nielsen Phys. Rev. A {\bf
66}, 032110 (2002).
%
\bibitem{bloc}
G. Vidal, J. I. Latorre, E. Rico, and A. Kitaev, Phys. Rev. Lett.
{\bf 90}, 227902 (2003).
%
\bibitem{zana}
H. T. Quan, Z. Song, X. F. Liu, P. Zanardi, and C. P. Sun, Phys.
Rev. Lett. {\bf 96}, 140604 (2006); P. Zanardi, H. T. Quan,
Xiaoguang Wang, and C. P. Sun, Phys. Rev. A {\bf 75}, 032109
(2007).
%
\bibitem{bologn}
L. Campos Venuti, C. Degli Esposti Boschi, M. Roncaglia, and A.
Scaramucci, Phys. Rev. A {\bf 73}, 010303 (2006); L. A. Wu, M. S.
Sarandy, and D. A. Lidar, Phys. Rev. Lett. {\bf 93}, 250404
(2004).
%
\bibitem{Hines}
A. P. Hines, G. J. Milburn and R. H. McKenzie, Phys. Rev. A
\textbf{71}, 042303 (2005).
%
\bibitem{emary}
C. Emary, N. Lambert, and T. Brandes, Phys. Rev. A 71, 062302
(2005).
%
\bibitem{brand}
N. Lambert, C. Emary, and T. Brandes, Phys. Rev. Lett. {\bf 92},
073602 (2004).
%
\bibitem{milburn}
A.P. Hines, C.M. Dawson, R.H. McKenzie and G.J. Milburn, Phys.
Rev. A {\bf 70}, 022303 (2004).
%
\bibitem{libertie}
G. Liberti, R. L. Zaffino, F. Plastina and F. Piperno, Phys. Rev.
A \textbf{73}, 032346 (2006).
%
\bibitem{vidal}
J. Vidal and S. Dusuel, Europhys. Lett. \textbf{74} 817 (2006).
%
\bibitem{libertif}
G. Liberti, F. Plastina and F. Piperno, Phys. Rev. A \textbf{74},
022324 (2006).
%
\bibitem{Jaynes}
E. T. Jaynes, F. W. Cummings, Proc. IEEE \textbf{51}, 89 (1963).
%
\bibitem{Tavis}
M. Tavis, F. W. Cummings, Phys. Rev. 170, \textbf{379} (1968).
%
\bibitem{dicke}
R. H. Dicke, Phys. Rev. \textbf{93}, 99 (1954).
%
\bibitem{leggett}
A. J. Leggett, S. Chakravarty, A. T. Dorsey, M. P. A. Fisher, A.
Gerg, and W. Zwerger, Rev. Mod. Phys. \textbf{59}, 1 (1987).
%
\bibitem{costi}
T.A. Costi and R.H. McKenzie, Phys. Rev. A {\bf 68}, 034301
(2003).
%
\bibitem{Longuet}
H. C. Longuet-Higgins, U. \"{O}pik, M. H. L. Pryce and R. A. Sack,
Proc. R. Soc. London, Ser. A \textbf{244}, 1 (1958).
%
\bibitem{eng}
R. Englman, \emph{The Jahn Teller Effect in Molecules and
Crystals} (Wiley, London, 1972) %
%
\bibitem{coffman}
V. Coffman, J. Kundu, and W. K. Wootters , Phys. Rev. A {\bf 61},
052306 (2000).
%
\bibitem{osborne2006}
T. J. Osborne and F. Verstraete, Phys. Rev. Lett. \textbf{96},
220503 (2006).
%
\bibitem{alteri}
K. A. Dennison and W. K. Wootters, Phys. Rev. A {\bf 65}, 010301
(2002); Y.-C. Ou and H. Fan, Phys. Rev.  A {\bf 75}, 062308
(2007);  M. Seevinck, Phys. Rev. A {\bf 76}, 012106 (2007).
%
\bibitem{firenze} T. Roscilde, P. Verrucchi, A. Fubini,
S. Haas, and V. Tognetti, Phys. Rev. Lett. {\bf 94}, 147208
(2005).
%
\bibitem{adesso}
G. Adesso, A. Serafini, and F. Illuminati, Phys. Rev. A {\bf 73},
032345 (2006); T. Hiroshima, G. Adesso, and F. Illuminati, Phys.
Rev. Lett. {\bf 98}, 050503 (2007).
%
\bibitem{mauro}
J. Lee, M. Paternostro, M. S. Kim, and S. Bose, Phys. Rev. Lett.
{\bf 96}, 080501 (2006); M. Paternostro, M. S. Kim, and G. M.
Palma, Phys. Rev. Lett. {\bf 98}, 140504 (2007); C. D. Ogden, M.
Paternostro, and M. S. Kim, Phys. Rev. A {\bf 75}, 042325 (2007).
%
\bibitem{tessier}
T. E. Tessier \emph{et al.}, Phys. Rev. A \textbf{68}, 062316
(2003).
%
\bibitem{osborne}
T. J. Osborne, Phys. Rev. A \textbf{72}, 022309 (2005).
%
\bibitem{bevilaqua}
G. Bevilacqua, L. Martinelli and G. Pastori Parravicini, Phys.
Rev. B \textbf{63}, 132403 (2000).
%
\bibitem{sato}
T. Sato, L. F. Chibotaru and A. Ceulemans, J. Chem. Phys.
\textbf{122}, 054104 (2005).
%
\bibitem{rungta}
P. Rungta, V. Buzek, C. M. Caves, H. Hillery and G. J. Milburn,
Phys. Rev. A \textbf{64}, 042315 (2001).
%
\bibitem{simon}
B. Simon and A. Dicke, Ann. Phys. {\bf 58}, \textbf{76} (1970).





\end{thebibliography}
\end{document}